\documentclass[pra, 10pt, twocolumn, floatfix, superscriptaddress]{revtex4-1}
\usepackage{amsmath} % need for subequations
\usepackage{graphicx} % need for figures
\usepackage{verbatim} % useful for program listings
\usepackage{color} % use if color is used in text
\usepackage{subfigure} % use for side-by-side figures
\usepackage{hyperref} % use for hypertext links, including those to external documents and
\usepackage{sidecap}
\usepackage{tikz}
\usepackage{amsmath,bm}
\usepackage{mathtools}
\usepackage{amsfonts}

 \DeclareMathSymbol{\minus}
       {\mathord}{operators}{"2D}

\begin{document}
%\begin{center}
%\rule[5ex]{\linewidth}{.1ex}
%\end{center}

\title{Creation and dynamics of two-dimensional skyrmions in antiferromagnetic spin-1 Bose-Einstein condensates}
\author{T. Ollikainen}
\author{E. Ruokokoski}
\affiliation{QCD Labs, COMP Centre of Excellence, Department of Applied Physics, Aalto University, P.O. Box 13500, FI-00076 AALTO, Finland}
\author{M. M\"ott\"onen$^{1,}$}
\affiliation{Low Temperature Laboratory, Aalto University, P.O.~Box 13500, FI-00076 AALTO, Finland}


\begin{abstract}
We numerically simulate the creation process of two-dimensional skyrmionic excitations in antiferromagnetic spin-1 Bose--Einstein condensates by solving the full three-dimensional dynamics of the system from the Gross--Pitaevskii equation. Our simulations reproduce quantitatively the experimental results of Choi \emph{et al.}, [Phys. Rev. Lett. {\bf 108}, 035301 (2012)] without any fitting parameters. Furthermore, we examine the stability of the skyrmion by computing the temporal evolution of the condensate in a harmonic potential. The presence of both the quadratic Zeeman effect and dissipation in the simulations is vital for reproducing the experimentally observed decay time.
\end{abstract}

\maketitle

\section{Introduction}

Optical trapping methods enable creation of Bose-Einstein condensates (BECs) with dynamics in the hyperfine spin degree of freedom~\cite{StamperKurn:1998, Kawaguchi2012253,RevModPhys.85.1191}. These so-called spinor BECs are described by a multi-component quantum field and they can host a wide variety of interesting topological excitations. Whereas the spectrum of topological excitations in scalar BECs with essentially no internal structure is rather limited, excitations such as coreless vortices~\cite{Mizushima:2001,Leanhardt:2003}, monopoles~\cite{Stoof:2001,PhysRevLett.91.190402,Pietila:2009}, and skyrmions~\cite{Khawaja:2001,PhysRevLett.91.010403,Leslie:2009} exist in spinor condensates. The imprinting of complex topological excitations in spinor BECs can be done in practice using time-dependent external magnetic fields~\cite{Isoshima:2000,Leanhardt:2002,Leanhardt:2003, PhysRevLett.99.250406,PhysRevLett.99.250406}.

  The concept of skyrmion originates from particle physics, where skyrmions were described as topological solitons in the nonlinear field theory for pions~\cite{Skyrme:1961}. The original Skyrme model accounts for $3+1$ dimensions and the nontrivial solutions to Skyrme's equations are three-dimensional (3D) skyrmions~\cite{PhysRevLett.86.3934}. Solitons in a modified Skyrme model with $2+1$ dimensions are referred to as two-dimensional (2D) skyrmions~\cite{Piette:1995}, which have later been studied in various fields of physics~\cite{Schmeller:1995,Xie:1996,Muhlbauer:2009,Fukuda:2011,Romming:2013}. We study the 2D skyrmions that occur as topological excitations in spinor BECs.

  Two-dimensional skyrmions were recently experimentally realized in antiferromagnetic BECs by Choi \emph{et al.}~\cite{Choi:2012}. The skyrmion creation process is based on ramping non-adiabatically a 3D quadrupole field through the condensate. In the adiabatic regime and ferromagnetic phase, a similar field ramp has been shown to generate a Dirac monopole~\cite{Pietila:2009, Ray:2014} or a multi-quantum vortex~\cite{Mottonen:2002,Leanhardt:2003, Leanhardt:2002, Kuopanportti:2013}. In Ref.~\cite{Choi:2012} the skyrmion was observed to decay into a uniform spin texture and it was speculated that the decay is due to the quadratic Zeeman shift and the induced spin currents. Recently, Huang \emph{et al.}~\cite{Huang:2013} simulated computationally the skyrmion creation process in a 2D system, and found that skyrmions with spiralling phase are formed. They also investigated the dynamics of the created skyrmions and observed no decay of the skyrmion even at long time scales. They speculate that the decay observed in the experiments~\cite{Choi:2012} may be caused by dissipation. Xu \emph{et al.}~\cite{Xu_Pra:2012} also investigated the skyrmion dynamics and observed that the skyrmion will decay due to dynamical mixing of the antiferromagnetic and ferromagnetic components. Due to their energy-conserving simulation, they did not observe the decay into uniform spin texture.

  We investigate the creation and stability of skyrmions in $^{23}$Na condensates with the aim at simulating accurately the experiments of Ref.~\cite{Choi:2012}. We numerically solve the full 3D dynamics of the mean-field spinor order parameter from the time-dependent Gross--Pitaevskii (f) equation. A very good quantitative agreement between the experiments and the simulations is achieved for the creation process without fitting parameters. We analyze the order parameter texture after the imprinting process and verify that it satisfies the skyrmion boundary conditions accurately. The dynamics of the created skyrmions are also considered and the effect of  dissipation and quadratic Zeeman term on the stability of the skyrmion is examined.

\section{Theory}

The dynamics of the mean-field order parameter $\Psi$ is solved from the time-dependent GP equation for spin-1 BEC. The effective Hamiltonian reads~\cite{Ohmi:1998,Ho}

\begin{equation}\label{eq:gp}
\begin{split}
\mathcal{H} &= -\frac{\hbar^2}{2m}\nabla^2 + V({\bf r })+ c_0 \Psi^\dagger\Psi \\&+ c_2 \Psi^\dagger {\bf F} \Psi \cdot {\bf F} + g_F \mu_B{\bf B}({\bf r},t)\cdot {\bf F} + q\left({\bf B}\cdot {\bf F}\right)^2,%include the Lagrange multiplier
\end{split}
\end{equation}
where $\hbar$ is the reduced Planck constant, $m$ is the mass of the constituent bosons, $V({\bf r})$ is the optical trapping potential, $g_F$ is the hyperfine Land\'e $g$-factor, $\mu_B$ is the Bohr magneton, ${\bf B}({\bf r},t)$ is the external magnetic field, ${\bf F} = (F_x, F_y, F_z)$ is the vector of the standard spin-1 matrices and $q$ is the quadratic Zeeman shift. For $^{23}$Na  $q=2\pi\hbar\times278\,\textrm{Hz}/\textrm{G}^2$ ~\cite{Ketterle:1998}. Here, $c_0$ and $c_2$ are coupling constants associated with the density--density and spin--spin interactions, respectively. The optical trapping potential is of the form $V(x,y,z) = \frac{1}{2}m\left(\omega_x^2x^2+\omega_y^2y^2+\omega_z^2z^2\right)$, where $\{\omega_k\}$ are the trapping frequencies in each spatial direction.

  The sign of the constant $c_2$ determines the natural magnetic phase of the condensate. For bare antiferromagnetic condensates, $c_2>0$, it is energetically favorable for the local spin to vanish. Condensates with $c_2<0$ are referred to as ferromagnetic, as they tend to maximize the local spin.

%It has been speculated that the reason for the instability observed in the experiments is dissipation, which has not been considered in simulations to date. In this work, the
The dissipation in the condensate is taken into account by introducing a single damping parameter in the master equation as
  \begin{equation}
  \label{eq:dissipation}
  i\hbar\partial_t\Psi({\bf r},t) = (1-i\Lambda) \mathcal{H}\Psi({\bf r},t),
  \end{equation}
  %where $\mathcal{H}^{\ast}=\mathcal{H}+p F_z$. The lagrange multiplier $p$ is introduced to account for the conservation of magnetization. 
where $\mathcal{H}$ is the Hamiltonian given in Eq.~(\ref{eq:gp}) and $\Lambda > 0$ is a dimensionless damping parameter which can be determined experimentally~\cite{Choi:1998}. The main dissipation channel in the system is the interaction of the condensate with the thermal cloud. Hence the damping parameter $\Lambda$ is expected to depend strongly on temperature. Note that Eq.~\eqref{eq:dissipation} is non-Hermitian and thus the norm of $\Psi$ is not conserved in the temporal evolution. Hence $\Psi$ is renormalized after each time step.
Choi et al.~\cite{Choi:1998} have shown that the dissipation observed in experiments~\cite{Jin:1996, Mewes:1996} is consistent with the solution of the GP equation with such a damping term and $\Lambda=0.03$. Furthermore, Tsubota et al.~\cite{Tsubota:2002} and  Kasamatsu et al.~\cite{Kasamatsu:2003} have used similar damping term to describe the effect of dissipation in the dynamics of a sudden rotated BEC in a trap and showed that the damping term successfully describes the vortex lattice formation process.

  The order parameter of a spin-1 BEC can be written in the form

  \begin{equation}
  \Psi = \left(\begin{array}{c}\psi_1 \\ \psi_0 \\ \psi_{-1}\end{array}\right) =  \sqrt{n}{\bf \zeta},
  \end{equation}
  where $n$ is the particle density and ${\bf \zeta}$ is a three-component spinor. The indices $1$, $0$, and $-1$ correspond to the eigenstates of $F_z$. The general form of the spinor for an antiferromagnetic condensate is given by~\cite{Ho}

  \begin{equation}
  \label{eq:spinor}
  {\bf \zeta} = e^{i\theta} U(\alpha,\beta,\gamma) \left(\begin{array}{c}0 \\ 1 \\ 0\end{array}\right) = e^{i\theta}\left( \begin{array}{c} -\frac{1}{\sqrt{2}}e^{-i\alpha}\sin \beta \\ \cos \beta \\ \frac{1}{\sqrt{2}}e^{i \alpha}\sin \beta \end{array} \right),
  \end{equation}
where $\alpha$, $\beta$ and $\gamma$ are the Euler angles, $U(\alpha,\beta,\gamma)=e^{-i F_z\alpha}e^{-i F_y\beta}e^{-i F_z\gamma}$ is the spin rotation operator, and $\theta$ is the 
scalar phase.

  In order to extract information on the magnetic ordering of the antiferromagnetic condensate, it is convenient to work in the Cartesian basis with the transformation~\cite{Mueller:2004}
  \begin{equation}
  \label{eq:cartesiantransform}
  \left( \begin{array}{c} \psi_x \\ \psi_y \\ \psi_z \end{array} \right) = \frac{1}{\sqrt{2}}\left( \begin{array}{ccc} -1 & 0 & 1 \\ -i & 0 & -i \\ 0 & \sqrt{2} & 0 \end{array} \right) \left( \begin{array}{c} \psi_1 \\ \psi_0 \\ \psi_{-1} \end{array} \right),
  \end{equation}
  where $(\psi_x,\psi_y,\psi_z)^T$ is the order parameter in the Cartesian basis. The magnetic ordering can be probed by the magnetic quadrupole moment matrix $Q$, the elements of which in the Cartesian basis are defined as~\cite{Mueller:2004}
  \begin{equation}
  Q_{ab} = \frac{\psi_a^*\psi_b^{ }+\psi_b^*\psi_a^{ }}{2|\psi|^2}.
  \end{equation}
  The unit vector $\hat{\bf{d}}$, corresponding to the largest eigenvalue of $Q$, characterizes the magnetic ordering. We refer to $\hat{\bf{d}}$ as the local magnetic axis. The pure antiferromagnetic order parameter in Eq.~\eqref{eq:spinor} can be written in the Cartesian basis as~\cite{Mueller:2004}
  \begin{equation}
  \Psi = \sqrt{n}e^{i\theta}\hat{\bf{d}},
  \end{equation}
where $n$ is the particle density. Note that a transformation $(\hat{\bf{d}},\theta)\to (-\hat{\bf{d}},\theta+\pi)$ has no effect on the order parameter. Thus $\hat{\bf{d}}$ is unoriented, though we illustrate it below as an oriented vector for clarity.
  In the effectively 2D system, the topological excitations related to the magnetic order are typically characterized by the topological charge
  \begin{equation}\label{eq:charge}
  Q_{\textrm{2D}} = \frac{1}{4\pi}\int dxdy \; \hat{\bf{d}}\cdot \left( \partial_x \hat{\bf{d}} \times \partial_y\hat{\bf{d}} \right).
  \end{equation}
  The topological charge describes the number of times the local magnetic axis in the $xy$-plane covers its configuration space, i.e., the unit sphere.
  Since the magnetic axis is unoriented, the charges $Q_{\textrm{2D}}$ and $-Q_{\textrm{2D}}$, defined in Eq.~(\ref{eq:charge}), arise from the same order parameter, and hence one can consider only non-negative charges.

  The local magnetic axis of an ideal 2D skyrmion is given in the cylindrical coordinates by~\cite{Piette:1995}
  \begin{equation}\label{eq:skyrmion}
  \hat{\bf{d}}(\rho,\phi) = \cos\!\beta(\rho)\,\hat{\bf{z}} + \sin\!\beta(\rho)\,\hat{\boldsymbol{\rho}},
  \end{equation}
  where $\beta (\rho)$ is a monotonically increasing function with boundary conditions $\beta(0) = 0$ and $\beta(\infty) = \pi$. The topological charge of the skyrmion texture is $Q_{\textrm{2D}} = 1$. For half-skyrmion the boundary conditions are $\beta(0)=0$ and $\beta(\infty) = \pi/2$ and the topological charge is $Q_{\textrm{2D}} = 1/2$.

  \section{Methods and parameters}
  \label{sc:methods}
  The external magnetic field is a combination of a 3D quadrupole field and a homogenous bias field ${\bf B}(\rho,z,t) = B'(t)\rho\hat{\boldsymbol{\rho}} + [B_z(t) - 2 B'(t)z]\hat{\bf{z}}$, where $B_z$ is the strength of the axial bias field and $B'$ is the radial magnetic field gradient.

In the beginning of the simulated skyrmion creation process, the field gradient is $B'(0)=8.1\,\textrm{G/cm}$ and the bias field strength is $B_z(0)=0.5$ G. Subsequently, $B_z$ is linearly ramped to value $B_z(T_1)=-0.5$ G in time $T_1$. We refer to this as the \emph{creation ramp}. The ramp time $T_1$ varies from $10$ ms to {84} ms. Immediately after the creation ramp the field gradient is linearly ramped down to zero in time $T_2=150\,\mu$s. Due to the nonadiabatic inversion of the bias field, the local magnetic axis $\hat{\bf{d}}$ of the condensate on the $z$-axis retain its original orientation and the tilt angle of $\hat{\bf{d}}$ with respect to $z$-axis increases with distance from the $z$-axis $\rho$. The tilt angle at the boundary of the condensate depends on the ramp rate and by varying $T_1$ we can generate boundary conditions corresponding to both skyrmions and half-skyrmions.

  In the experiment, the condensate is created in the absence of the quadrupole field and before the creation ramp, the quadrupole field is adiabatically ramped on. This operation has no significant effect on the order parameter as verified by Huang \emph{et al.}~\cite{Huang:2013}. Hence, in the simulations, we solve the initial state of the condensate from the GP equation by a relaxation method with conditions $B'(0)=8.1\,\textrm{G/cm}$, $B_z(0)=0.5\,\textrm{G}$ and requiring that the local spin vanishes.

   The simulation parameters were chosen to match the experimental values used by Choi \emph{et al.}~\cite{Choi:2012}. The optically trapped $^{23}$Na BEC consists of $N=1.2\times10^{6}$ atoms and the optical trapping frequencies are $(\omega_x,\omega_y,\omega_z) = 2\pi \times (3.5, 4.6, 430)$ Hz. The hyperfine Land\'e $g$-factor for $^{23}$Na is $g_F = -1/2$ and the interaction strengths in Eq.~\eqref{eq:gp} are $c_0=(g_0+2g_3)/3$ and $c_2=(g_2-g_0)/3$, where $g_0 \approx 46 \cdot \frac{4\pi \hbar^2 a_B}{m}$, $g_2 \approx 52 \cdot \frac{4\pi \hbar^2 a_B}{m}$, and $a_B$ is the Bohr radius~\cite{Ho}.

The split operator method together with fast Fourier transformations are utilized in the computation of the temporal evolution. The set of equations received from Eq.~(\ref{eq:gp}) are numerically solved in a discretized three-dimensional grid of size $200 \times 200 \times 30$, with the total volume of approximately $336 \times 336 \times 8$ $\mu$m$^3$. Time step of $\tau$ = 0.45 $\mu$s is used.
We did not simulate the expansion of the condensate. Choi \emph{et al.}~\cite{Choi:2012} observed that the condensate expansion in the transverse direction is less that 10\% and hence the effect of expansion on the $z$-integrated density profiles is small. 

  \begin{figure}[t]
  \centering
  \includegraphics[width=\linewidth,keepaspectratio]{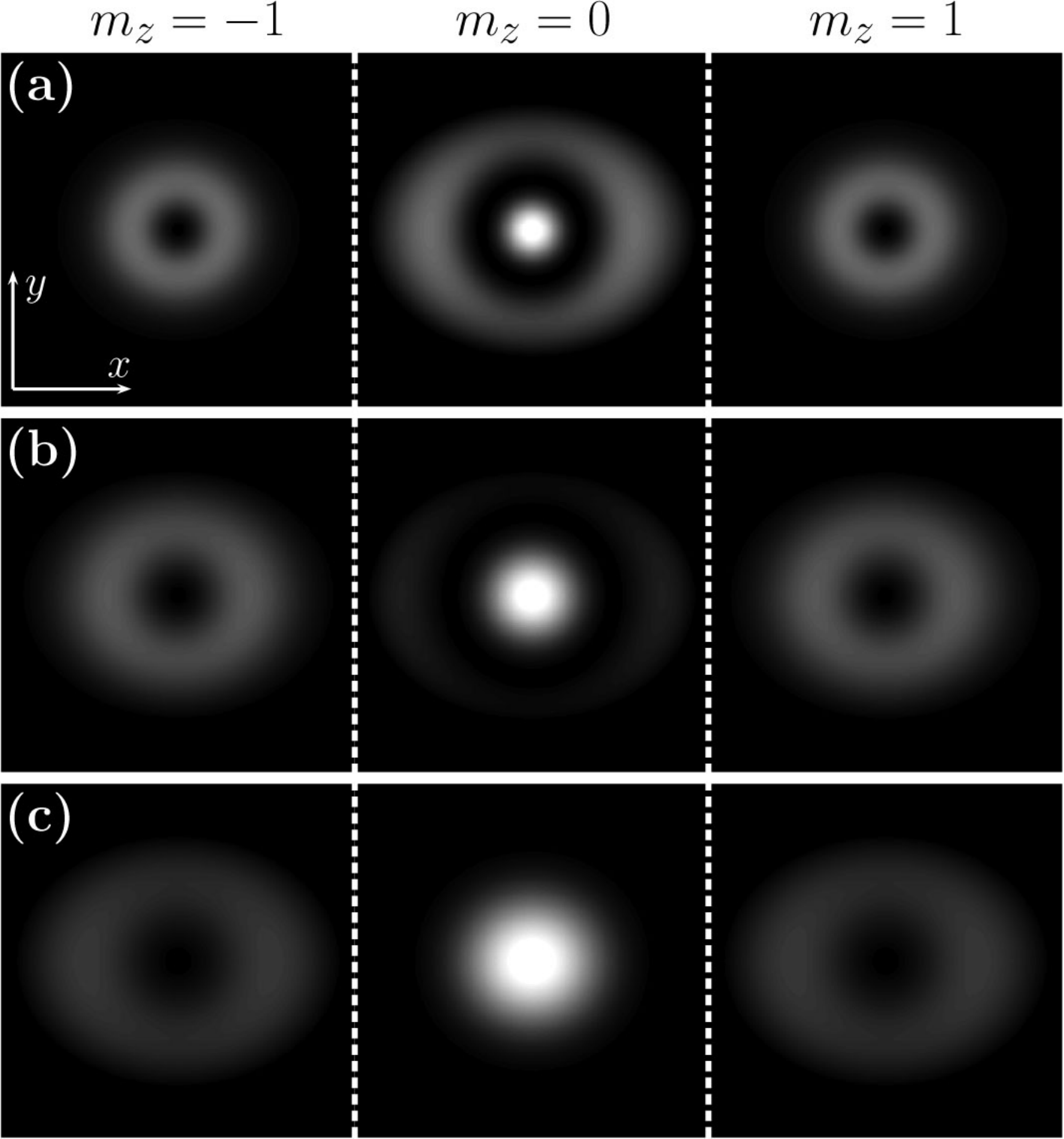}
  \caption{Integrated particle densities of the created skyrmions. The three columns correspond to the $z$-quantized $\psi_0$ and $\psi_{\pm1}$ spinor components as indicated. The components in each row are separated for clarity, but in the simulations, their origins overlap. The densities are scaled such that white shows the peak density decreasing linearly to zero denoted by black color. The panels correspond to bias field ramp rates (a) $|\dot{B}_z| =$ 12 G/ms, (b) 32 G/ms and (c) 80 G/ms. The field of view in each panel is $300 \times 300$ $\mu$m$^2$. See Sec.~\ref{sc:methods} for other parameter values.}
  \label{fig:tiheys}
  \end{figure}

  \begin{figure}[t]
  \centering
  \includegraphics[width=\linewidth,keepaspectratio]{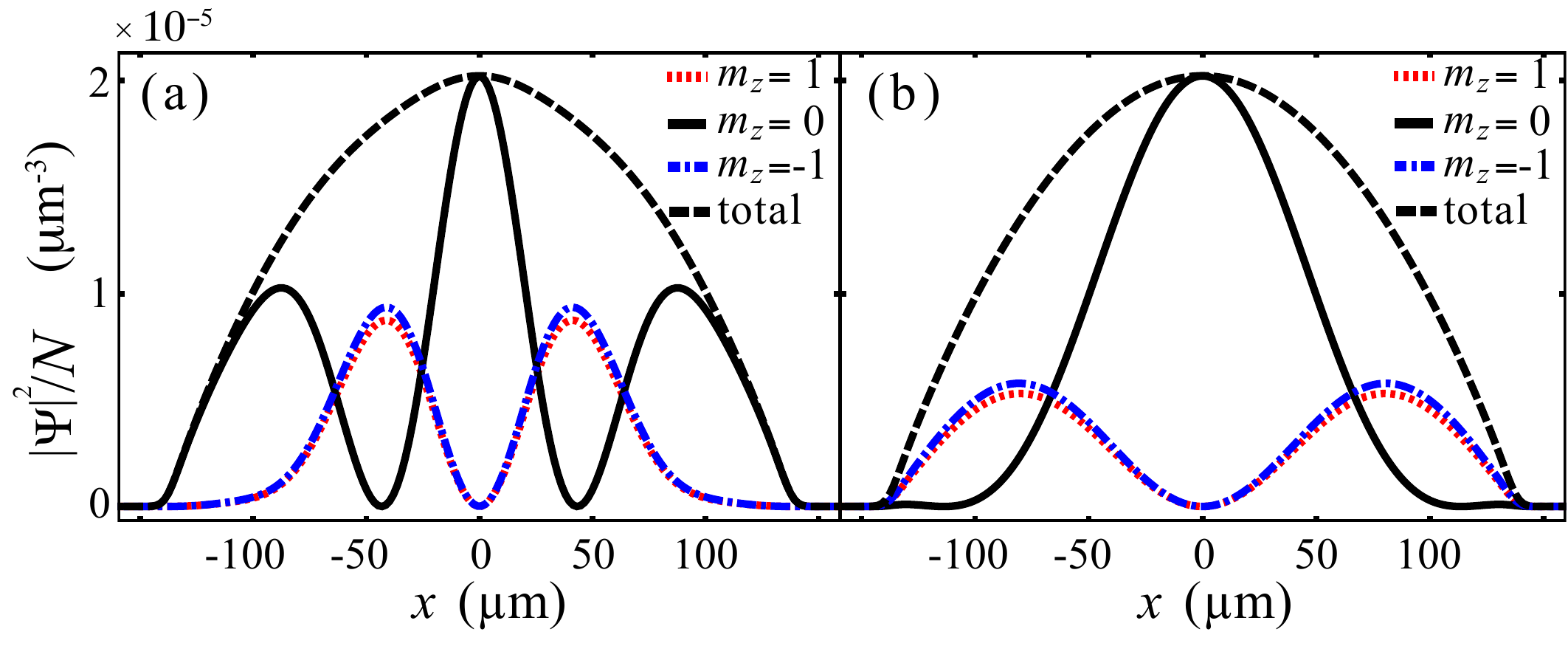}
  \caption{Particle density distributions in different spin components together with the total density traced along $x$-axis of the created (a) skyrmion and (b) half-skyrmion. The dotted red line corresponds to the $m_z=1$ component, the solid black line to the $m_z=0$ component, the dash-dotted blue line to the $m_z=-1$ component, and the dashed black line to the total density. For panel (a) $|\dot{B}_z|$ = 12 G/ms and for panel (b) 80 G/ms. The rest of the parameters are given in Sec. III.}
  \label{fig:skyrmionkuva}
  \end{figure}

  \section{Results}
\subsection{Creation}
  
  The density distributions of the condensate after the skyrmion creation process described in Sec.~\ref{sc:methods} are presented in Fig.~\ref{fig:tiheys}. The densities are integrated along the $z$-axis consistent with the imaging method employed in the experiments of Ref.~\cite{Choi:2012}. Horizontal traces of the particle density along the $x$-axis after the imprinting process are shown in Fig.~\ref{fig:skyrmionkuva}. The size of the generated texture can be characterized by the radius $R_{\pi/2}$ of the density-depleted ring in the $m_z=0$ component. The ring indicates the location where $\beta = \pi/2$. The dependence of the radius $R_{\pi/2}$ on the azimuthal angle is negligible in the $xy$-plane since the size of the texture is determined by the external magnetic field which is radially symmetric. The fact that the optical trapping frequencies differ in the $x$- and $y$-directions affects the aspect ratio of the total particle density but has no significant effect on the locations of the density extrema of individual spinor components.

\begin{figure}[t]
  \centering
  \includegraphics[width=\linewidth,keepaspectratio]{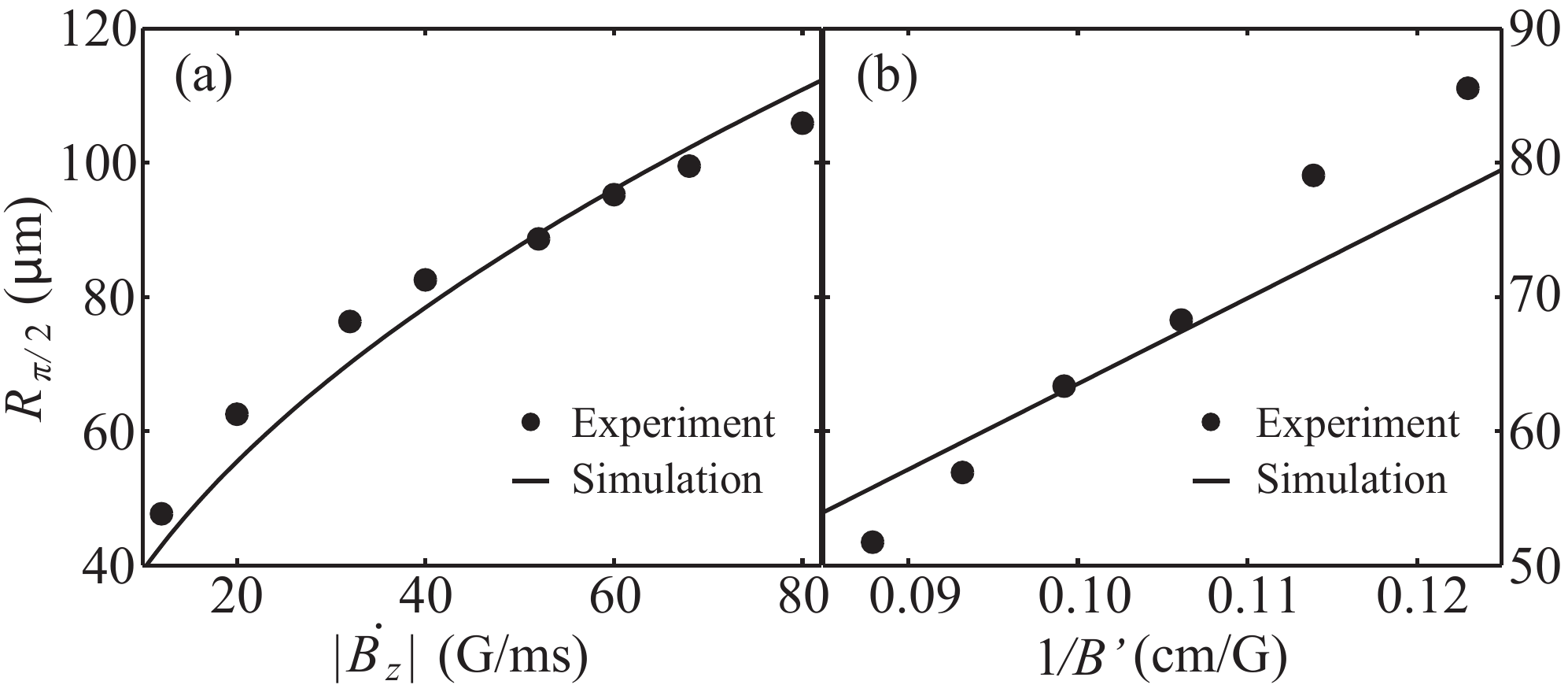}
  \caption{Size of the skyrmion $R_{\pi/2}$ as a function of (a) the magnitude of the ramp rate of the bias field $| \dot{B}_z |$ with $B' = 8.1$ G/cm, and (b) the inverse of the radial field gradient $1/B'$ with $|\dot{B}_z| = 40$ G/ms. The solid lines represent the simulation results, and the dots correspond to the experimental values obtained from Ref.~\cite{Choi:2012}. A square-root function and linear function were fitted to the simulated results with essentially no deviation. Thus the simulated points are not shown for clarity. The fitting functions were (a) $R_{\pi/2} = 12.4\sqrt{\dot{B}_z}\times10^{-7}\,$s$^{1/2}$G$^{-1/2}$m and (b) $R_{\pi/2} = 638\times 1/B'\times 10^{-4}\,$G$-0.314\times 10^{-6}$ m.}
  \label{fig:grkuva}
  \end{figure}

  It was found in the previous studies~\cite{Choi:2012,Huang:2013} that the size of the created skyrmion increases with the speed of the magnetic-field inversion $|\dot{B}_z|$. The dependence of the radius $R_{\pi/2}$ on the ramp rate of the bias field $|\dot{B}_z|$ and on the strength of the quadrupole field $B'$ is given in Fig.~\ref{fig:grkuva}. The simulated results are in very good quantitative agreement with the experiment~\cite{Choi:2012}. We attribute the possibly remaining small discrepancy of the numerical and experimental results to small differences in the parameter values employed in the simulations compared to those actually present in the experiments. In the experiments, the density distributions are measured after expansion but this yields a qualitatively different correction to Fig.~\ref{fig:grkuva} from what is required to achieve a complete agreement between the numerical and experimental results. Furthermore, the relative radial expansion is reported in Ref.~\cite{Choi:2012} to be less than 10\%.

  \begin{figure}[t]
  \centering
  \includegraphics[width=\linewidth,keepaspectratio]{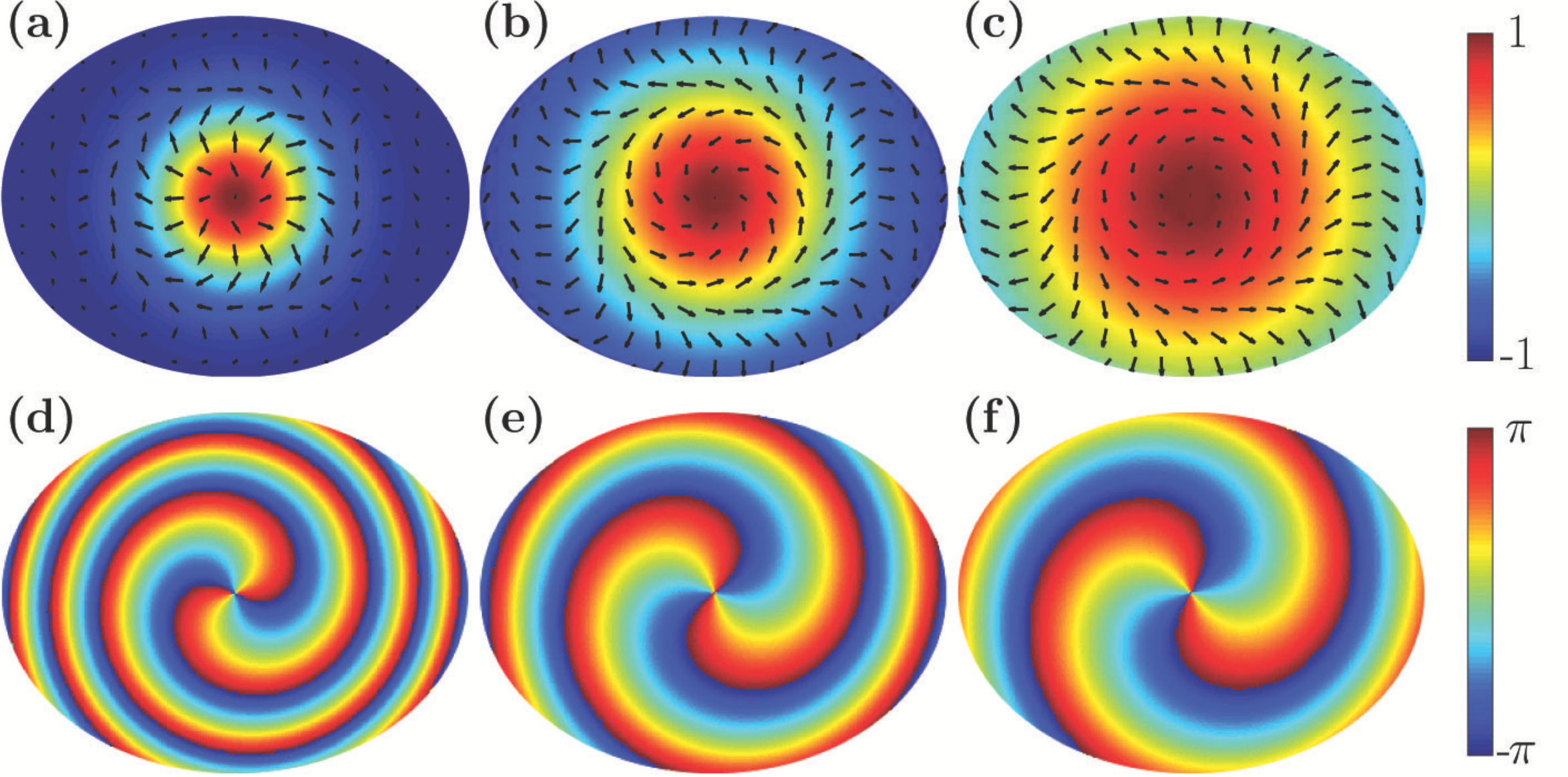}
  \caption{Direction of the local magnetic axis (a)--(c) and the phase of $\psi_1/\psi_{-1}$ (d)--(f) right after the bias field $B_z$ is inverted and the quadrupole field $B'$ is ramped down. The ramp rate is $|\dot{B}_z|$ = 12 G/ms in panels (a) and (d), $|\dot{B}_z|$ = 32 G/ms in (b) and (e), and $|\dot{B}_z|$ = 80 G/ms in (c) and (f). The field gradient is $B'$ = 8.1 G/cm. In panels (a)--(c) the arrows represent the projection of the vector $\hat{\bf{d}}$ to the $xy$-plane and the $z$-component is presented with the colormap. The top colormap corresponds to the value of $d_z$, and the bottom one to the phase. Regions with particle density $|\Psi|^2 \le 10^{-10} $N$/\mu $m$^3$ are colored white.}
  \label{fig:phase}
  \end{figure}

  The direction of the local magnetic axis of the created skyrmion
  is presented in Figs.~\ref{fig:phase}(a)--\ref{fig:phase}(c). The spin dynamics are two-dimensional in the condensate, since the spin healing length is greater than the thickness of the cloud~\cite{Choi:2012}. In Figs.~\ref{fig:phase}(a) and \ref{fig:phase}(b), the skyrmion boundary conditions are met, as the function $\beta(\rho)$ continuously changes from $\beta(0)=0$ to $\beta[R(x,\,y)]=\pi$, where $R(x,\,y)$ is the spatial extent of the condensate in the $xy$-plane. This is depicted by the fact that the local magnetic axis continuously changes its orientation from $\hat{\bf{d}}(0)=\hat{\bf{z}}$ to $\hat{\bf{d}}(R)=-\hat{\bf{z}}$. For a half-skyrmion in Fig.~\ref{fig:phase}(c), the angle is continuously tilted from $\beta(0)=0$ to $\beta[R(x,\,y)]=\pi/2$, as $\hat{\bf{d}}(0)$ = $\hat{\bf{z}}$ and $d_z[R(x,\,y)] = 0$.

  The apparent presence of the azimuthal component of $\hat{\bf d} $ and the spiraling phases in the $\psi_{\pm 1}$  spinor components [see Figs.~\ref{fig:phase}(d)--\ref{fig:phase}(f)] indicate that the created texture is not an ideal skyrmion described by Eq.~(\ref{eq:skyrmion}). The spiraling phases are caused by the population of breathing modes due to the fast inversion of the bias field in the presence of the quadrupole field. This phenomenon can also be described by the phase acquired in the spatially dependent Landau--Zener process. The created magnetic texture can be characterized by introducing a function $\gamma(\rho)$ in Eq.~(\ref{eq:skyrmion}), such that $\hat{\bf{d}}(\rho,\phi) = \cos \beta(\rho)\hat{\bf{z}} + \sin\beta(\rho)\left[\cos\!\gamma(\rho)\,\hat{\boldsymbol{\rho}} + \sin\!\gamma(\rho)\,\hat{\boldsymbol{\phi}}\right]$~\cite{Choi_Njop:2012}. The spiral shape is also observed in Ref.~\cite{Huang:2013}. The slower the inversion, the tighter the spiral, since we are working far in the nonadiabatic regime.

  \begin{figure}[t]
  \centering
  \includegraphics[width=\linewidth,keepaspectratio]{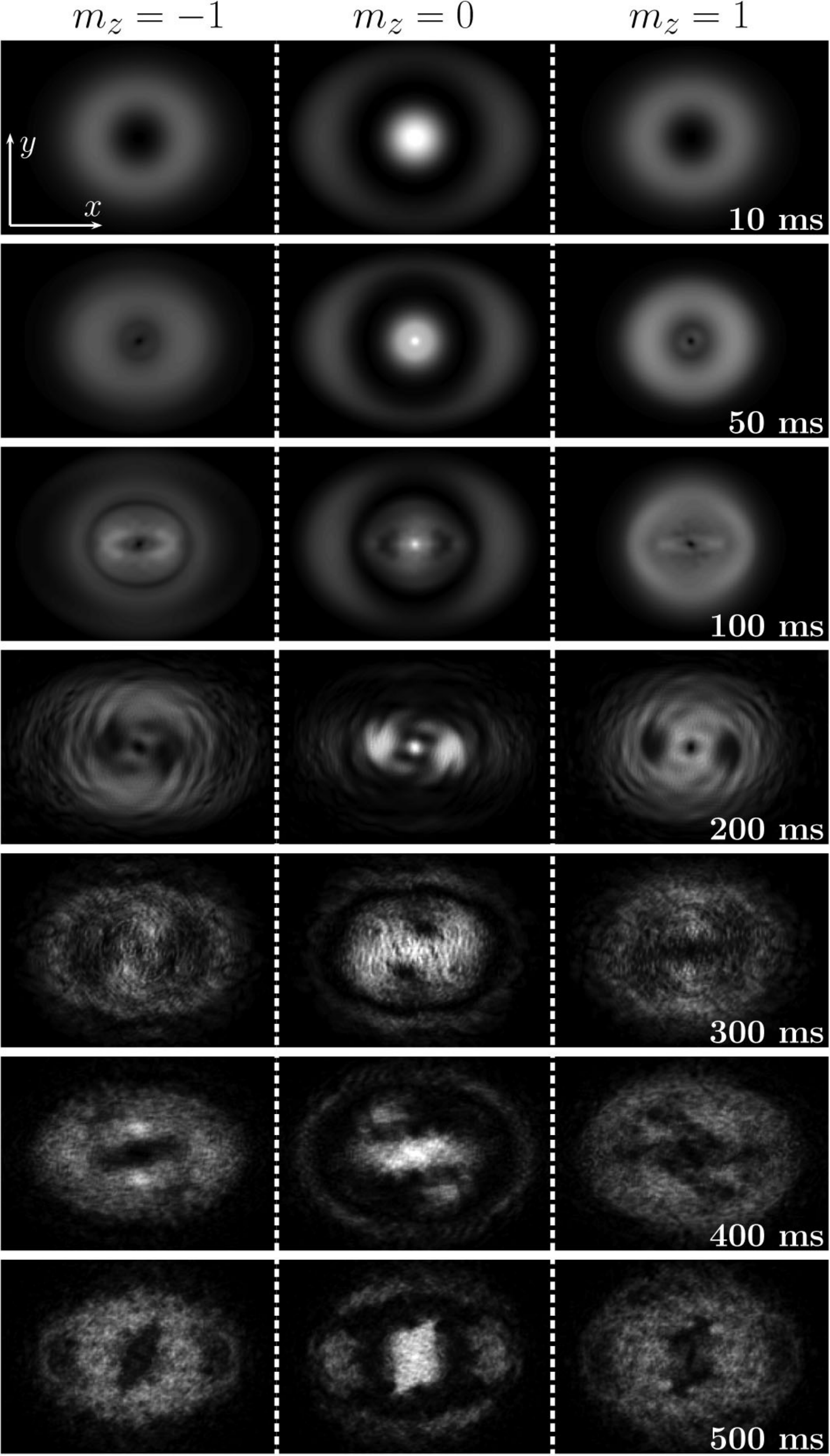}
  \caption{Temporal evolution of the skyrmion in a harmonic optical potential. The skyrmion was created with parameters $|\dot{B}_z|$ = 20 G/ms and $B'$ = 8.1 G/cm. Dissipation is not taken into account. The field of view for the $z$-integrated particle densities in each panel is $300 \times 210$ $\mu$m$^2$.}
  \label{fig:deformation}
  \end{figure}

  \begin{figure}[t]
  \centering
  \includegraphics[width=\linewidth,keepaspectratio]{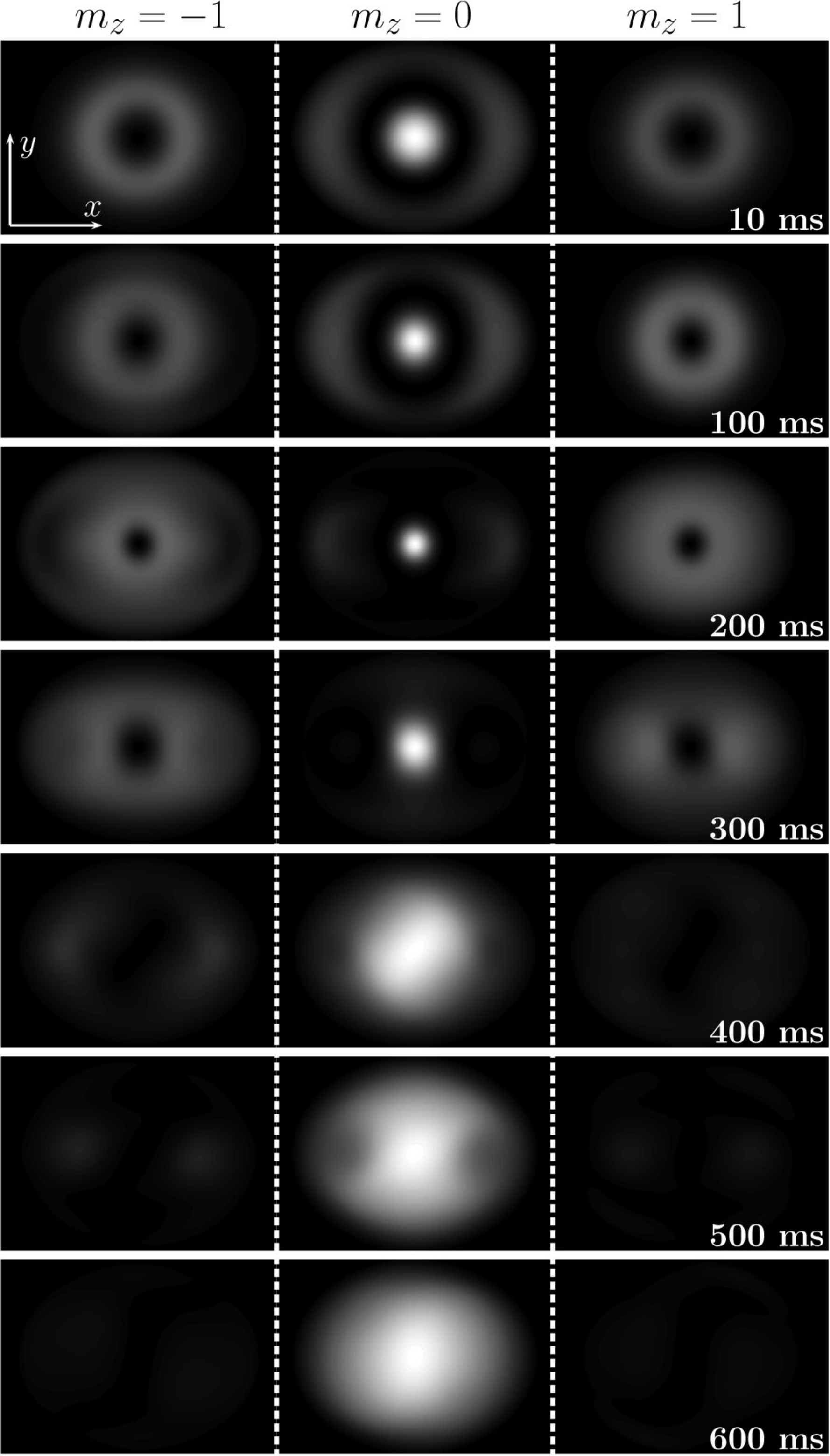}
  \caption{Temporal evolution of the skyrmion in a harmonic optical potential without external magnetic fields. The damping factor is set to $\Lambda = 0.05$. The skyrmion is created with parameters $|\dot{B}_z|$ = 20 G/ms and $B'$ = 8.1 G/cm. The field of view for the $z$-integrated particle densities in each panel is $300 \times 210$ $\mu$m$^2$.}
  \label{fig:dissipation}
  \end{figure}

\subsection{Dynamics}
  The dynamics of the created skyrmionic state was studied for multiple values of the dissipation parameter $\Lambda$ and for the quadratic Zeeman effect stregths $q=0$ and $q=2\pi\hbar\times278\,\textrm{HzG}^{-2}$. The optical trap was not altered and the bias field was at value $B_z=-0.5\,\textrm{G}$ throughout. The axial magnetization of the created skyrmion is zero. Hence, due to the conservation of magnetization, the linear Zeeman term in the Hamiltonian (\ref{eq:gp}) has no other effect on the dynamics, than causing the spin of the condensate atoms to precess in the $xy$-plane with high Larmor frequency~\cite{Ketterle:1998, Saito:2005} To avoid these rapid oscillations, we set the linear Zeeman term to zero here. 

The dynamics in the absence of the quadratic Zeeman term and dissipation ($q=\Lambda=0$) is shown in Fig.~\ref{fig:deformation}.  We observe breathing of the condensate, and excitation of surface modes is clearly visible after $200$ ms of decay dynamics. The created texture is destroyed, but some skyrmion-like properties remain in the condensate, i.e., the depleted density ring of the $m_z = 0$ component seems to be present in Fig.~\ref{fig:deformation} even after $500\,\textrm{ms}$, and the $m_z = \pm 1$ components occupy the depleted area. The fraction of particles in the $m_z=\pm1$ components is not decreasing during the temporal evolution. 

It has been suggested that an antiferromagnetic order parameter with the skyrmion texture will evolve into a mixture of both ferromagnetic and antiferromagnetic domains~\cite{Xu_Pra:2012,Huang:2013}. This behavior is also present in Fig.~\ref{fig:deformation}; for example, at $400$ ms, the $m_z=-1$ component has two density maxima in regions where the $m_z=1$ component is depleted. This indicates that some ferromagnetic domains are present in the condensate.

In order to investigate the effect of dissipation, we set the damping parameter in Eq.~\eqref{eq:dissipation} to $\Lambda > 0$. The dissipative evolution with $\Lambda = 0.05$ and $q=0$  is shown in Fig.~\ref{fig:dissipation}. During the temporal evolution the $m_z = \pm 1$ components disappear and the condensate evolves into a texture with essentially all the atoms in the $m_z = 0$ component. At 600 ms only the $m_z=0$ component has significant particle density. The decay time is longer than that observed in the experiments~\cite{Choi:2012} although $\Lambda=0.05$ corresponds to rather strong dissipation that is not expected to be present in the experiments due to low temperature.

\begin{figure}[t]
\centering
\includegraphics[width=\linewidth,keepaspectratio]{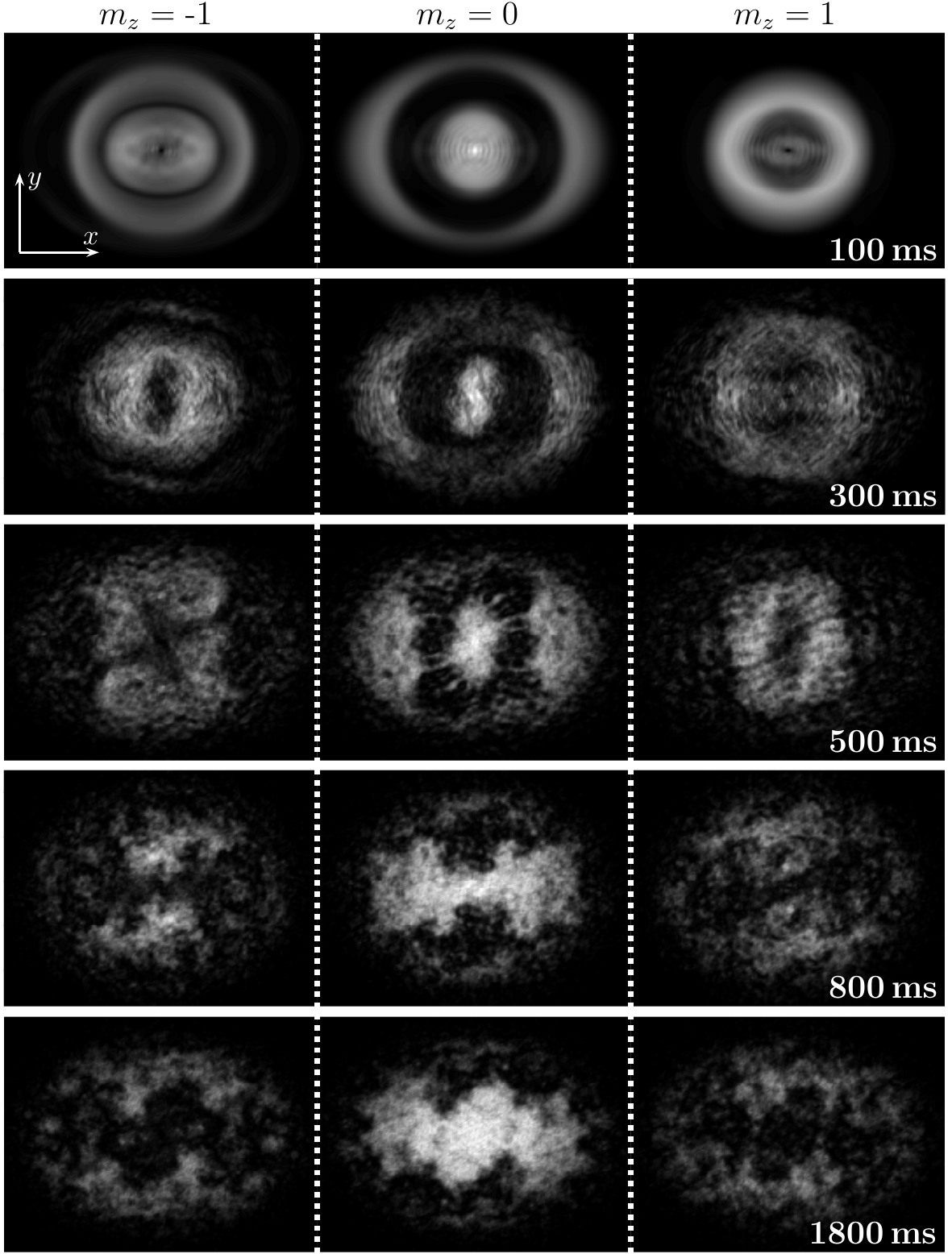}
\caption{\label{fig:000}Temporal evolution of the skyrmion in a harmonic optical potential with $\Lambda=0$ and $q=2\pi\hbar\times278\,\textrm{Hz}/\textrm{G}^2$. The skyrmion is created with parameters $|\dot{B}_z|$ = 20 G/ms and $B'$ = 8.1 G/cm. The field of view for the $z$-integrated particle densities in each panel is $300 \times 210$ $\mu$m$^2$.}
\end{figure}

The temporal evolution of the skyrmion defect in the presence of the quadratic Zeeman shift $q=2\pi\hbar\times278\,\textrm{Hz}/\textrm{G}^2$ and in the absence of dissipation $\Lambda = 0$, is shown in Fig.~\ref{fig:000}. The skyrmion decays and the fraction of particles in the $m_z=\pm1$ states is diminished. The decay is, however, very slow and even after $1.8\,\textrm{s}$ only $52$\% of the particles reside in the $m_z=0$ state. In the experiments~\cite{Choi:2012}, the uniform texture in the $m_z=0$ state was reached already at $t\le300\,\textrm{ms}$.
\begin{figure}
\centering
\includegraphics[width=\linewidth,keepaspectratio]{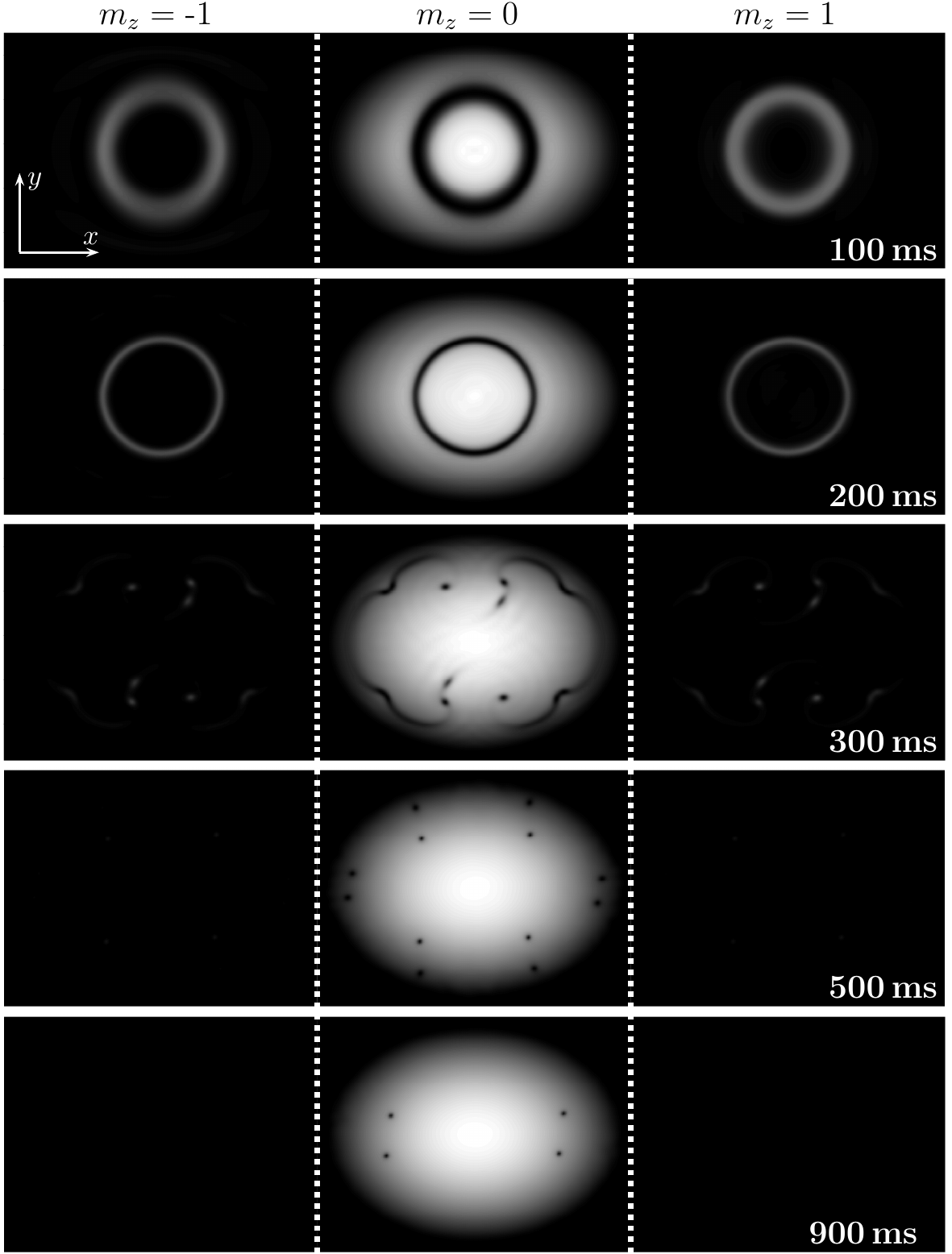}
\caption{\label{fig:0015}Temporal evolution of the skyrmion in a harmonic optical potential with $\Lambda=0.015, q=2\pi\hbar\times278\,\textrm{Hz}/\textrm{G}^2$. The skyrmion is created with parameters $|\dot{B}_z|$ = 20 G/ms and $B'$ = 8.1 G/cm. The field of view for the $z$-integrated particle densities in each panel is $300 \times 210$ $\mu$m$^2$.}
\end{figure}

If dissipation is included together with the quadratic Zeeman effect, the skyrmion decay becomes faster. For small values of $\Lambda$, the decay process is quantitatively similar to that in Fig.~\ref{fig:000}, but the uniform texture in the $m_z=0$ component is reached faster. For $\Lambda\geq0.005$, the skyrmion decay process changes dramatically featuring the formation of half-quantum vortex-antivortex pairs. The temporal evolution for $\Lambda=0.015$ is shown in Fig.~\ref{fig:0015}. The $m_z=\pm1$ components occupy the density-depleted regions of the $m_z=0$ state as was observed in the experiments ~\cite{Choi:2012}. Some of the half-quantum vortices annihilate shortly after their formation, but some persist in the condensate and holes are developed in the total particle density as the fraction of particles in the $m_z=\pm 1$ states is decreased in the decay process. Slowly the remaining vortices move towards the boundary of the condensate, but even for $t=900$ ms some vortices remain. We also investigated larger values of $\Lambda$ and the results were qualitatively similar to those in Fig.~\ref{fig:0015}, except that the full occupation of the $m_z=0$ state was reached faster. With $\Lambda=0.015$ the rate at which the full occupation of the $m_z=0$ state is reached approximately matches the experiments~\cite{Choi:2012}.

We also investigated the effect of Gaussian noise in the initial order parameter on the dynamics, but it did not change the qualitative features of the decay process or speed up the decay.
\subsection{Interference}
\begin{figure}
%\vspace{10pt}
\centering
\includegraphics[width=\linewidth,keepaspectratio]{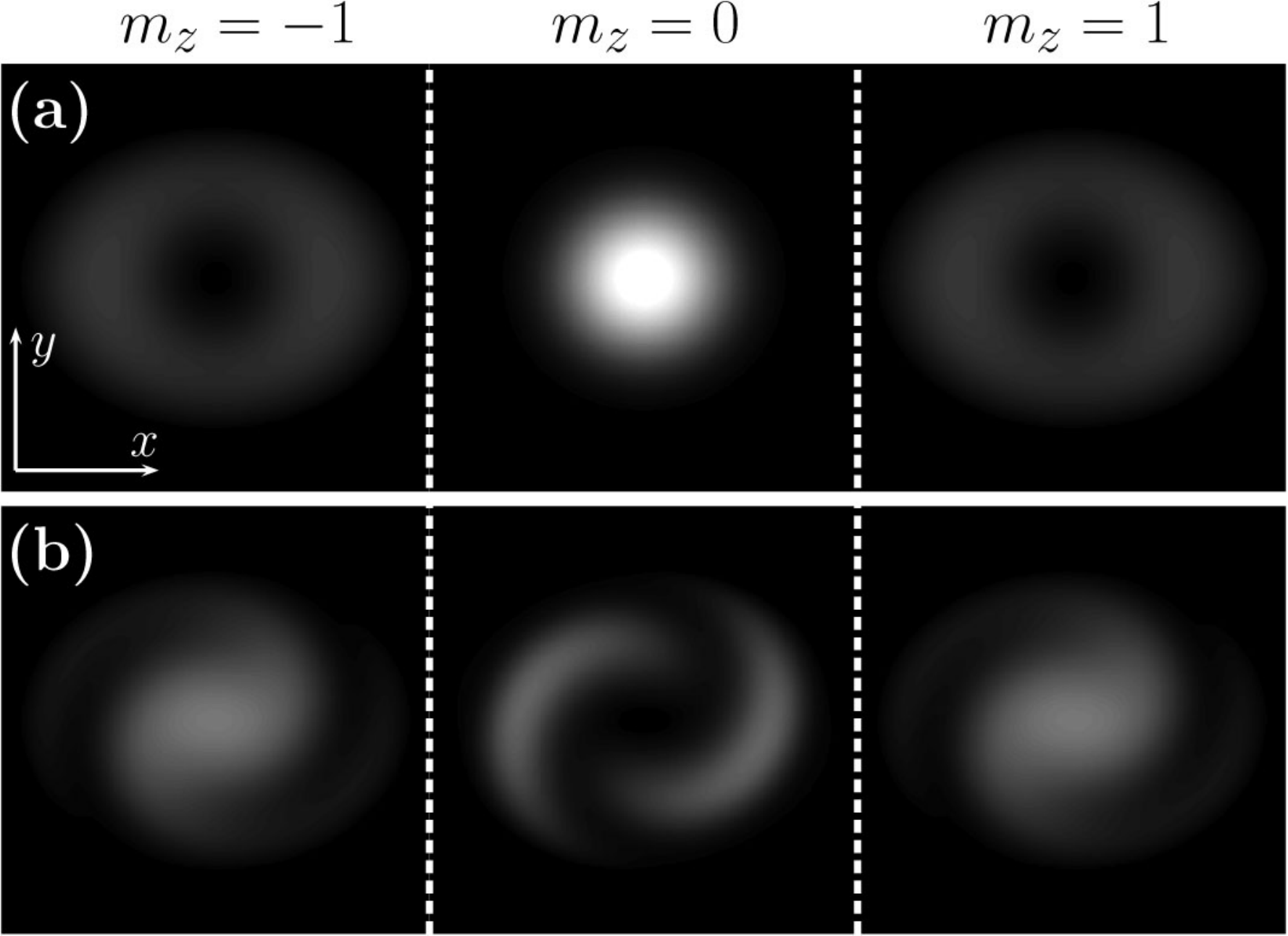}
\caption{Interference pattern of the half-skyrmion with a $\pi/2$-pulse along $y$-axis applied after the creation ramp. The initial density distribution just after the creation ramp is shown in row (a) and the resulting interference pattern in row (b). Here, $|\dot{B}_z| = 80$ G/ms and $B' = 8.1$ G/cm in the skyrmion creation process. The field of view in each panel showing the $z$-integrated particle density is $300 \times 300$ $\mu$m$^2$.}
\label{fig:pulse}
\end{figure}

  We investigate the interference patterns obtained by rotating the local magnetic axis $\hat{\bf{d}}$ by $\pi/2$ about the $y$- or $x$-axis. These rotations correspond to the effect of $y-\pi/2$ or $x-\pi/2$ pulse on the half-skyrmion state. The application of a $\pi/2$ pulse in the $x$- or $y$-direction transfers atoms from $m_z = 0$ state equally to the $m_z = {\pm 1}$ states. We find in our simulations that there is a relative phase difference of 4$\pi$ between the phases of the $m_z = \pm 1$ components, depicted by the phase observed in Fig.~\ref{fig:phase}. The spiral shape in the phase causes the transferred atoms to form a crescent shape as shown in Fig.~\ref{fig:pulse}. Similar interference patterns have been observed previously with skyrmions and vortices in ferromagnetic BECs~\cite{Leslie:2009,Wright:2009}.

\section{Conclusions}
We have studied the creation process and dynamics of skyrmion textures in the local magnetic axis of spin-1 BECs. The parameters of the creation process were chosen to match those of the experimental setup of Ref.~\cite{Choi:2012}. It was found that 2D skyrmions are indeed created, and they feature spiraling phases. Both the magnitude of the radial magnetic field and the field inversion rate have an effect on the size of the skyrmion texture and on the tightness of the spirals in the phase structure. The spiraling phase causes the atoms to form crescent-shape density distributions if a spin rotation corresponding to the experimental rf pulse is applied.

  During the temporal evolution of the condensate in a harmonic potential, the created skyrmion excitations are destroyed. Without dissipation and quadratic Zeeman term, the fraction of particles in the $m_z=\pm 1$ components is not decreased. Furthermore, the density rings in $m_z = \pm1$ components are present even half a second after the skyrmion has been created. Thus it can be argued that some skyrmion-like properties remain in the condensate. We find that the initially antiferromagnetic order parameter develops some ferromagnetic features as has been suggested in recent theoretical studies~\cite{Xu_Pra:2012,Huang:2013}. If the quadratic Zeeman term and dissipation are included, the skyrmion decays into half-quantum vortex-antivortex pairs and the decay times observed in the experiment are reached. However, some vortices remain in the condensate even after $900\,\textrm{ms}$. We conclude that neither the dissipation nor the quadratic Zeeman term alone is enough to cause the skyrmion to decay in time scale observed in the experiment, but their combination is sufficient. The fact that  we observe vortices even $900$ ms after the creation ramp although no density depletions are observed in the experiment after $300\, \textrm{ms}$, can be due to the  simplistic way of including the dissipation. In the employed model for dissipation, the relative decay rate for all the excitations is given by $\Lambda$ in the linear regime. We fix this parameter to yield a matching time scale of the spinor decay dynamics into the $m_z=0$ component between the simulation and the experiment. Thus it is natural that the time scale of the vortex decay dynamics influenced by the thermal gas trapped in the vortex core is not simultaneously captured by this model.

\begin{acknowledgments}
We acknowledge the financial support by Academy of Finland through its Centre of Excellence Program (Grant No. 251748) and Grants No. 135794, No. 272806, and No. 141015, and by the Finnish Doctoral Program in Computational Sciences. We thank CSC -- IT Center for Science Ltd. and Aalto Science-IT project for providing computational resources.
\end{acknowledgments}
\bibliography{skyrmion}
\bibliographystyle{apsrev4-1}

\end{document}